\documentclass[11pt]{article}%
\usepackage{amsmath}
\usepackage{amssymb}
\usepackage{amsfonts}

\usepackage{cite}
\usepackage{graphicx}
\usepackage{float}
\usepackage{longtable}
\usepackage[utf8]{inputenc}
\usepackage{hyperref}

\usepackage{algorithmicx}
\usepackage[ruled,vlined]{algorithm2e}
\SetAlgoCaptionSeparator{ }

\usepackage{fullpage}

\newcommand{\fref}[1]{Fig.~\ref{#1}}

\newcommand{\sref}[1]{Section~\ref{#1}}

\newenvironment{algo}[1][!h]
  {
   \begin{algorithm}[#1]%
  }{\end{algorithm}}

\providecommand{\U}[1]{\protect\rule{.1in}{.1in}}

\begin{document}

\title{Entanglement Accessibility Measures for the Quantum Internet}
\author{Laszlo Gyongyosi\thanks{School of Electronics and Computer Science, University of Southampton, Southampton SO17 1BJ, U.K., and Department of Networked Systems and Services, Budapest University of Technology and Economics, 1117 Budapest, Hungary, and MTA-BME Information Systems Research Group, Hungarian Academy of Sciences, 1051 Budapest, Hungary.}
\thanks{Parts of this work were presented in \cite{qkdrev}.}
\and Sandor Imre\thanks{Department of Networked Systems and Services, Budapest University of Technology and Economics, 1117 Budapest, Hungary.}
}

\date{}

\maketitle
\begin{abstract}
We define metrics and measures to characterize the ratio of accessible quantum entanglement for complex network failures in the quantum Internet. A complex network failure models a situation in the quantum Internet in which a set of quantum nodes and a set of entangled connections become unavailable. A complex failure can cover a quantum memory failure, a physical link failure, an eavesdropping activity, or any other random physical failure scenario. Here, we define the terms entanglement accessibility ratio, cumulative probability of entanglement accessibility ratio, probabilistic reduction of entanglement accessibility ratio, domain entanglement accessibility ratio, and occurrence coefficient. The proposed methods can be applied to an arbitrary topology quantum network to extract relevant statistics and to handle the quantum network failure scenarios in the quantum Internet.
\end{abstract}

\section{Introduction}
\label{sec1}
As quantum computers evolves significantly \cite{qc1,qc2,qc3,qc4,qc5,qc6,qcadd1,qcadd2,qcadd3,qcadd4}, there arises a fundamental need for a communication network that provides unconditionally secure communication and all the network functions of the traditional internet. This network structure is the quantum Internet \cite{ref1,ref2,ref3, refn7,puj2,puj1}. The availability of quantum entanglement is a crucial aspect in any global-scale quantum Internet. The quantum Internet refers to a set of connected heterogeneous quantum communication networks realized by quantum nodes and channels (such as optical fibers or wireless optical quantum channels in the physical layer) \cite{ref23,ref24,ref25,ref26,ref27,ref11,ref28}. The quantum Internet also integrates a set of classical auxiliary communication channels to transmit auxiliary classical side-information between the quantum nodes. The quantum Internet is modeled as a global-scale quantum communication network composed of quantum subnetworks and networking components.
The core network of the quantum Internet is assumed to be an entangled network structure \cite{ref1,ref5,ref6,nadd1,nadd2,nadd3,nadd4,nadd5,nadd6,nadd7,nadd8,nadd9,netadd1,netadd2,uj1,uj2}, which is a communication network in which the quantum nodes are connected by entangled connections. An entangled connection refers to a shared entangled system (i.e., a Bell state for qubit systems to connect two quantum nodes) between the quantum nodes. In an unentangled network structure, the quantum nodes are not necessarily connected by entanglement \cite{ref11,ref38}, and the communication between the nodes is realized in a point-to-point setting. This setting does not allow quantum communication over arbitrary distances, and an unentangled network structure can mostly be used for establishing a point-to-point quantum key distribution (QKD) \cite{qkdrev,ref30}  between the quantum nodes. These short distances can be extended to longer distances by the utilization of free-space quantum channels \cite{ref1,ref30}. However, this solution is auxiliary, since it can be used only at some specific points of the unentangled network structure. Therefore, it does not represent an adequate and fundamental answer to the problem of long-distance quantum communication. Consequently, in an unentangled network structure, the multi-hop settings are weak for experimental, long-distance and global-scale quantum communication. On the other hand, the entangled network structure allows the parties to establish multi-hop entanglement, multi-hop QKD, high-precision sensor networks, advanced distributed computations and cryptographic functions, advanced quantum protocols, and, more importantly, the distribution of quantum entanglement over arbitrary (unlimited, in theory) distances \cite{ref1}. As an important corollary, an entangled network structure provides a strong experimental basis for realizing a global-scale quantum communication network, the quantum Internet.

In the entangled network structure of the quantum Internet, the entangled connections form entangled paths. Entanglement between a distant source and a target node is established through several intermediate repeater nodes \cite{ref1,ref5,ref6,ref7,ref13}. The level of entanglement (i.e., the level of an entangled connection) is defined as the number of nodes (i.e., the hop-distance between entangled nodes) spanned by the shared entanglement, whose range is extended by the basic operation of entanglement swapping (entanglement extension). The entangled connections have several relevant attributes, the most important of which are the fidelity of entanglement and the entanglement throughput. The throughput of an entangled connection is measured as the number of entangled states per second at a given fidelity, which provides a useful metric on the basis of which further relevant metrics can be built.

Here we define measures to characterize the ratio of accessible quantum entanglement in case of complex network failures \cite{ref14,ref15,ref16,ref17} in the quantum Internet. A \textit{complex network failure} models a network situation in which a set of quantum nodes and a set of entangled connections become unavailable because of an (unknown) reason. A complex failure, therefore, can cover a set of practical failure reasons: a quantum memory failure situation in which a set of nodes and connections become unavailable, quantum node and connection failure scenarios, physical-link failures or an eavesdropping activity. Specifically, a complex failure event is modeled by a network domain that is referred to as a complex failure domain. In our model, a failure domain has an abstracted center point and a given length radius \cite{ref16,ref17}. This domain approach allows us to describe the probability that a given node or entangled connection (i.e., a given network element) is affected by a failure in the function of the given network element's distance from the abstracted center point of the complex failure domain. 

The \textit{entanglement accessibility ratio} of a given quantum network is based on the metric of the given entangled connection's entanglement throughput. Each entangled connection is further verified by a given condition that puts a lower bound on the entanglement throughput. The entanglement accessibility ratio measures the successful accessible entanglement at a given lower bound condition for parallel complex failures in the quantum network. 

We also define the \textit{cumulative probability of entanglement accessibility ratio} that quantifies the cumulative probability of all complex failure events' occurrence for which the entanglement accessibility ratio exceeds a given lower bound. 

We also quantify the probability that the total entanglement accessibility ratio in the quantum network is reduced to at most a particular ratio after a complex failure. Particularly, this parameter is referred as the \textit{probabilistic reduction of entanglement accessibility ratio}. 

To describe the impacts of a given complex failure on the ratio of accessible entanglement, we define the \textit{domain entanglement accessibility ratio}, which quantifies the accessible entanglement ratio after a complex failure in a particular domain in a function of the radius of the given failure domain. 

We define the \textit{occurrence coefficient of an entanglement accessibility ratio} (occurrence ratio) at a complex failure domain, which is measured by the ratio of the number of occurrence of a given entanglement accessibility ratio in the network after a complex failure event and the total number of occurrences of all entanglement accessibility ratios after a complex failure event. 

We show that the defined measures can be extracted from the occurrence ratio, and therefore, it is enough to determine the occurrence coefficient to derive the other metrics. We propose an algorithm to determine the occurrence coefficient from the empirical quantities of the quantum network that are directly observable in the analyzed network setting. In particular, the defined entanglement accessibility measures can be derived in a purely empirical way by extracting relevant statistics from the analyzed quantum network. 

The proposed protocol is not dependent from the actual physical implementation, therefore it can be applied in the heterogeneous network structure and network components of the quantum Internet (the protocol can also be applied in the quantum Internet at the utilization of magnetic field in the perturbation method \cite{pert0,pert1,pert2} (kind of Zeeman Effect \cite{zeem}) in the physical layer \footnote{At a constant magnetic field perturbation, the evolution operator is diagonal. Even when the magnetic field depends only on time and not on space, the exact perturbation unitary evolution operator remains diagonal. The quantum system can be disturbed by perturbing it with electric, magnetic or electromagnetic radiation and hence, the system becomes excited and changes its state. Magnetic field-based protocol design here is more complex because a 1-D (dimensional) magnetic field will not act on a 1-D charged particle (From the Lorentz Law \cite{lor}, $\left( qV\times B \right)\bot V$, where $q$ is the charge of the particle, $V$ is the velocity, and $B$ is the magnetic field.). Note, a charged particle can also be excited in a 3-D box with a 3-D control magnetic field. Another possible extension to this problem is to consider a particle in a 3-D box perturbed by a vector electric field and a vector magnetic field \cite{em1,em2}.}, or in electromagnetic field-based \cite{em1,em2} scenarios in the network components.).

The novel contributions of our manuscript are as follows:

\begin{enumerate}
\item  We define measures to characterize the accessible quantum entanglement in case of complex network failures in the quantum Internet.

\item  We define the terms entanglement accessibility ratio, cumulative probability of entanglement accessibility ratio, probabilistic reduction of entanglement accessibility ratio, and occurrence coefficient.

\item  We show that the defined measures can be extracted from the occurrence ratio, and therefore, it is enough to determine the occurrence coefficient to derive the other metrics. 

\item  We propose an algorithm to determine the occurrence coefficient from the empirical quantities of the quantum network that are directly observable in the analyzed network setting of the quantum Internet.

\item  The entanglement accessibility measures can be derived in a purely empirical way by extracting relevant statistics from the quantum Internet.
\end{enumerate}

This paper is organized as follows. In \sref{relw}, the related works are summarized. In \sref{sec2}, some preliminaries are introduced. \sref{sec3} defines the entanglement accessibility measures. \sref{sec4} discusses the occurrence coefficient and defines an algorithm for the empirical evaluation of the measures. In \sref{nume}, a numerical evaluation is proposed. Finally, \sref{sec5} concludes the paper.

\section{Related Works}
\label{relw}
In this section, we review some recent results connected to the establishment of the experimental quantum Internet. 

A technical roadmap on the experimental development of the quantum Internet has been provided in \cite{refn7}. The roadmap is connected to the Quantum Internet Research Group (QIRG) \cite{refqirg}, which group is formulated and supported by an international researcher background and collaboration. The authors of \cite{refn7} address some important capability milestones for the realization of a global-scale quantum Internet. The technical roadmap also addresses important future engineering problems brought up by the quantum Internet, such as the development of a standardized architectural framework for the quantum Internet, standardization and protocols of the quantum Internet, application programming interface (API) for the quantum Internet, and the definition of the application level of the quantum Internet \cite{qkdrev}.

In a quantum Internet scenario, entanglement purification is a procedure that takes two imperfect systems $\sigma _{1} $ and $\sigma _{2} $ with initial fidelity $F_0<1$, and outputs a higher-fidelity density $\rho $ such that $F\left(\rho \right)>F_0$. In \cite{refn5}, the authors propose novel physical approaches to assess and optimize entanglement purification schemes. The proposed solutions provide an optimization framework of practical entanglement purification.

In \cite{refn3}, the authors defined a method for deterministic delivery of quantum entanglement on a quantum network. The results allow us to realize entanglement distribution across multiple remote quantum nodes in a quantum Internet setting. 

In \cite{sat}, a satellite-to-ground QKD system over 1,200 kilometres has been demonstrated. The proposed model integrated a low-Earth-orbit satellite with decoy-state QKD. The reported key rate of the protocol was above the kHz key rate over a distance up to 1200 km. The work has a relevance for an experimental quantum Internet, since the results also allow us to realize high-efficiency long-distance QKD in a global quantum Internet setting.

In \cite{telep}, the authors demonstrated the quantum teleportation of independent single-photon qubits over 1,400 kilometres. Since an experimental realization of a global-scale quantum Internet requires the application of quantum teleportation over long-distances, the proposed results represent a fundamental of any experimental quantum Internet. In \cite{refn4}, the authors demonstrated quantum teleportation with high fidelity values between remote single-atom quantum memories. 

Some other recent results connected to the development of an experimental global-scale quantum Internet are as follows. In \cite{refn1}, the authors demonstrated the Bell inequality violation using electron spins separated by 1.3 kilometres. In \cite{refn2}, the authors demonstrated modular entanglement of atomic qubits using photons and phonons. The quantum repeaters are fundamental networking elements of any experimental quantum Internet. The quantum repeaters are used in the entanglement distribution process to generate quantum entanglement between distant senders and receivers. The quantum repeaters also realize the entanglement purification and the entanglement swapping (entanglement extension) procedures. For an experimental realization of quantum repeaters based on atomic ensembles and linear optics, see \cite{refn6}.

Since quantum channels also have a fundamental role in the quantum Internet, we suggest the review paper of \cite{ref4}, and also the work of \cite{ref12}, for some specialized applications of quantum channels. For a review on some recent results of quantum computing technology, we suggest \cite{add4}. For some recent services developed for the quantum Internet, we suggest \cite{ref8,ref9,ref10,add1,add2,add3,ref19,ref20,ref21,poisson}.

Some other related topics are as follows. The works \cite{ref4,ref5,ref6,ref7,ref8,ref9,ref10,ref11} are related to the utilization of entanglement for long-distance quantum communications and for a global-scale quantum Internet, and also to the various aspects of quantum networks in a quantum Internet setting. 

For some fundamental works on quantum machine learning, see \cite{ref31,ref32,ref45,ref60}, on quantum Shannon theory, see \cite{ref4,ref12,ref22,ref29,ref34,ref46,ref61}, on quantum computing see \cite{ref33,ref35}, for schemes for reducing decoherence in quantum memory see \cite{ref36}, for quantum network coding see \cite{ref40,ref41,ref42,ref62}, for transformation of multipartite pure states, see \cite{ref43}, for multistage entanglement swapping see \cite{ref55}, while for optical microcavities and photonic channels for quantum communication, see \cite{ref47}. 

For some important works on the experimental implementations of quantum repeaters, entanglement purification and entanglement distribution, see \cite{ref37,ref39,ref44,ref48,ref49,ref50,ref51,ref52,ref53,ref54,ref55,ref56,ref57,ref58,ref59}.

\section{Preliminaries}
\label{sec2}
\subsection{Entanglement Fidelity}
The aim of the entanglement distribution procedure is to establish a $d$-dimensional entangled system between the distant points $A$ and $B$, through the intermediate quantum repeater nodes. Let $d=2$, and let ${\left| \beta _{00}  \right\rangle} $ be the target entangled system $A$ and $B$, ${\left| \beta _{00}  \right\rangle} =\frac{1}{\sqrt{2} } \left({\left| 00 \right\rangle} +{\left| 11 \right\rangle} \right),$ subject to be generated. At a particular density $\sigma $ generated between $A$ and $B$, the fidelity of $\sigma $ is evaluated as
\begin{equation} \label{ZEqnNum728497} 
F=\left\langle  {{\beta }_{00}} | \sigma |{{\beta }_{00}} \right\rangle .
\end{equation} 
Without loss of generality, an aim of a practical entanglement distribution is to reach $F\ge 0.98$ in \eqref{ZEqnNum728497} for a given $\sigma $ \cite{ref1,ref2,ref3,ref4,ref5,ref6,ref7,ref8}.

\subsection{Entangled Network Structure}
Let $V$ refer to the nodes of an entangled quantum network $N$, which consists of a transmitter node $A\in V$, a receiver node $B\in V$, and quantum repeater nodes $R_{i} \in V$, $i=1,\ldots ,q$. Let $E=\left\{E_{j} \right\}$, $j=1,\ldots ,m$ refer to a set of edges (an edge refers to an entangled connection in a graph representation) between the nodes of $V$, where each $E_{j} $ identifies an ${\rm L}_{l} $-level entanglement, $l=1,\ldots ,r$, between quantum nodes $x_{j} $ and $y_{j} $ of edge $E_{j} $, respectively. Let $N=\left(V,{\rm {\mathcal S}}\right)$ be an actual quantum network with $\left|V\right|$ nodes and a set ${\rm {\mathcal S}}$ of entangled connections. An ${\rm L}_{l} $-level, $l=1,\ldots ,r$, entangled connection $E_{{\rm L}_{l} } \left(x,y\right)$, refers to the shared entanglement between a source node $x$ and a target node $y$, with hop-distance 
\begin{equation} \label{ZEqnNum202142} 
d\left(x,y\right)_{{\rm L}_{l} } =2^{l-1} ,                                               
\end{equation} 
since the entanglement swapping (extension) procedure doubles the span of the entangled pair in each step. This architecture is also referred to as the doubling architecture \cite{ref1,ref5,ref6,ref7}. 

For a particular ${\rm L}_{l} $-level entangled connection $E_{{\rm L}_{l} } \left(x,y\right)$ with hop-distance \eqref{ZEqnNum202142}, there are $d\left(x,y\right)_{{\rm L}_{l} } -1$ intermediate nodes between the quantum nodes $x$ and $y$.

\subsection{Entanglement Purification and Entanglement Throughput}
Entanglement purification is a probabilistic procedure that creates a higher fidelity entangled system from two low-fidelity Bell states. The entanglement purification procedure yields a Bell state with an increased entanglement fidelity $F'$, 
\begin{equation} \label{3)} 
F_{in} <F' \le 1,                                                 
\end{equation} 
where $F_{in} $ is the fidelity of the imperfect input Bell pairs. The purification requires the use of two-way classical communications \cite{ref1,ref2,ref3,ref4,ref5,ref6,ref7,ref8}.

Let $B_{F} (E_{{\rm L}_{l} }^{i})$ refer to the entanglement throughput of a given ${\rm L}_{l} $ entangled connection $E_{{\rm L}_{l} }^{i} $ measured in the number of $d$-dimensional entangled states established over $E_{{\rm L}_{l} }^{i} $ per sec at a particular fidelity $F$ (dimension of a qubit system is $d=2$) \cite{ref1,ref2,ref3,ref4,ref5,ref6,ref7,ref8}. 

For any entangled connection $E_{{\rm L}_{l} }^{i} $, a condition $c$ should be satisfied, as 

\begin{equation} \label{ZEqnNum801212}
c:{{B}_{F}}( E_{{{\text{L}}_{l}}}^{i})\ge {B}_{F}^{\text{*}}( E_{{{\text{L}}_{l}}}^{i}),\text{ for }\forall i,
\end{equation} 
where ${{B}}_{F}^{\text{*}}( E_{{{\text{L}}_{l}}}^{i})$ is a critical lower bound on the entanglement throughput at a particular fidelity $F$ of a given $E_{{{\text{L}}_{l}}}^{i}$, i.e., ${{B}_{F}}( E_{{{\text{L}}_{l}}}^{i})$ of a particular $E_{{{\text{L}}_{l}}}^{i}$ has to be at least ${B}_{F}^{\text{*}}( E_{{{\text{L}}_{l}}}^{i})$.

\section{Model Description}
\label{sec3}
In this section, we define the terms and metrics for entanglement accessibility in the quantum Internet.

\subsection{Failure Identifications in the Quantum Internet}

Let ${\rm {\mathcal R}}_{f} $ refer to a complex failure domain that models a set of quantum nodes $V\left({\rm {\mathcal R}}_{f} \right)$ and a set of entangled connections ${\rm {\mathcal S}}\left({\rm {\mathcal R}}_{f} \right)$ in a particular network domain \cite{ref16,ref17}, whose nodes and entangled connections are affected by a complex failure $f$ (complex -- randomly affects both nodes and connections). Note, that while ${\rm {\mathcal S}}\left({\rm {\mathcal R}}_{f} \right)$ refers to the set of local entangled connections within the failure domain ${\rm {\mathcal R}}_{f}$, set $E$ refers to the entangled connections of the global quantum network $N$, therefore $\mathcal{S}\left( {{\mathcal{R}}_{f}} \right)$ is a subset of $E$, 
\begin{equation}
\mathcal{S}\left( {{\mathcal{R}}_{f}} \right)\subset E, 
\end{equation}
and
\begin{equation}
V\left( {{\mathcal{R}}_{f}} \right)\subset V, 
\end{equation}
also holds.

An $f$ complex failure event is identified by the entanglement throughput of an $i$-th ${\rm L}_{l} $-level entangled connection $E_{{\rm L}_{l} }^{i} $ as
\begin{equation} \label{5)} 
f:{{B}_{F}}( E_{{{\text{L}}_{l}}}^{i})<{B}_{F}^{\text{*}}( E_{{{\text{L}}_{l}}}^{i}),
\end{equation} 
where ${B}_{F}^{\text{*}}( E_{{{\text{L}}_{l}}}^{i})$ is a critical lower bound on the entanglement throughput.

In the $c_{{\rm {\mathcal R}}_{f} } $ center of ${\rm {\mathcal R}}_{f} $, for all entangled connections of the set ${\rm {\mathcal S}}\left({\rm {\mathcal R}}_{f} \right)$ of ${\rm {\mathcal R}}_{f} $,
\begin{equation} \label{6)} 
B_{F} (E_{{\rm L}_{l} }^{i})=0,                                                          
\end{equation} 
and therefore, the probability $\Pr \left(f\right)$ that an event $f$ occurs at $c_{{\rm {\mathcal R}}_{f} } $ for all elements of ${\rm {\mathcal S}}\left({\rm {\mathcal R}}_{f} \right)$ is
\begin{equation} \label{7)} 
\Pr \left(f\right)=1.                                                           
\end{equation} 

As the distance $d$ from the center of ${\rm {\mathcal R}}_{f} $ increases, the complex failure probability $\Pr \left(f\right)$ decreases, e.g., 
\begin{equation} \label{ZEqnNum668817} 
\Pr \left(f\right)<1.                                                           
\end{equation} 

Let $c_{{\rm {\mathcal R}}_{f} } $ be the center of domain ${\rm {\mathcal R}}_{f} $, and let $r_{{\rm {\mathcal R}}_{f} } $ be the radius of ${\rm {\mathcal R}}_{f} $
defined as in terms of the hop-distance of an abstracted shortest entangled path $\mathcal{P}$ in ${\rm {\mathcal R}}_{f} $, as
\begin{equation} \label{ZEqnNum483007} 
r_{{\rm {\mathcal R}}_{f} } =d\left(\mathcal{P}\left(x\left(c_{{\rm {\mathcal R}}_{f} } \right),y\left(c_{{\rm {\mathcal R}}_{f} } \right)\right)\right), 
\end{equation} 
where $x\left(c_{{\rm {\mathcal R}}_{f} } \right)\in {\rm {\mathcal R}}_{f} $ is the nearest affected quantum node to $c_{{\rm {\mathcal R}}_{f} } $, $y\left(c_{{\rm {\mathcal R}}_{f} } \right)\in {\rm {\mathcal R}}_{f} $ is the farthest affected quantum  node from $c_{{\rm {\mathcal R}}_{f} } $, while $\mathcal{P}\left(x\left(c_{{\rm {\mathcal R}}_{f} } \right),y\left(c_{{\rm {\mathcal R}}_{f} } \right)\right)$ is an abstracted shortest entangled path between $x\left(c_{{\rm {\mathcal R}}_{f} } \right)$ and $y\left(c_{{\rm {\mathcal R}}_{f} } \right)$, with a hop-distance $d\left({\rm {\mathcal P}}\left(x\left(c_{{\rm {\mathcal R}}_{f} } \right),y\left(c_{{\rm {\mathcal R}}_{f} } \right)\right)\right)$, as
\begin{equation} \label{eq2)} 
\begin{split}
   d\left( \mathcal{P}\left( x\left( {{c}_{{{\mathcal{R}}_{f}}}} \right),y\left( {{c}_{{{\mathcal{R}}_{f}}}} \right) \right) \right) &\\ 
  =&d{{\left( x\left( {{c}_{{{\mathcal{R}}_{f}}}} \right),{{{{x}'_{1}}}}\left( {{c}_{{{\mathcal{R}}_{f}}}} \right) \right)}_{{{\text{L}}_{l}}}}+\sum\limits_{i=1}^{m}{d{{\left( {{{{x}'_{i}}}}\left( {{c}_{{{\mathcal{R}}_{f}}}} \right),{{{{x}'_{i+1}}}}\left( {{c}_{{{\mathcal{R}}_{f}}}} \right) \right)}_{{{\text{L}}_{l}}}}} \\ 
  &+d{{\left( {{{{x}'_{m+1}}}}\left( {{c}_{{{\mathcal{R}}_{f}}}} \right),y\left( {{c}_{{{\mathcal{R}}_{f}}}} \right) \right)}_{{{\text{L}}_{l}}}},  
\end{split}
\end{equation} 
where $x'_{i} \left(c_{{\rm {\mathcal R}}_{f} } \right)$, $i=1,\ldots ,m$ are intermediate quantum nodes between $x\left(c_{{\rm {\mathcal R}}_{f} } \right)$ and $y\left(c_{{\rm {\mathcal R}}_{f} } \right)$ on the entangled path ${\rm {\mathcal P}}\left(x\left(c_{{\rm {\mathcal R}}_{f} } \right),y\left(c_{{\rm {\mathcal R}}_{f} } \right)\right)$. 

Thus, \eqref{ZEqnNum483007} can be rewritten via \eqref{eq2)}. Then, assuming a doubling architecture on $\mathcal{P}\left(x\left(c_{{\rm {\mathcal R}}_{f} } \right),y\left(c_{{\rm {\mathcal R}}_{f} } \right)\right)$ between $x\left(c_{{\rm {\mathcal R}}_{f} } \right)$ and $y\left(c_{{\rm {\mathcal R}}_{f} } \right)$ in ${\rm {\mathcal R}}_{f}$, the radius in \eqref{ZEqnNum483007} is yielded as
\begin{equation} \label{ZEqnNum384071} 
\begin{split}
  {{r}_{{{\mathcal{R}}_{f}}}}&\\ 
 & ={{2}^{l\left( E\left( x\left( {{c}_{{{\mathcal{R}}_{f}}}} \right),{{{{x}'_{1}}}}\left( {{c}_{{{\mathcal{R}}_{f}}}} \right) \right) \right)-1}}+\sum\limits_{i=1}^{m}{{{2}^{l\left( E\left( {{{{x}'_{i}}}}\left( {{c}_{{{\mathcal{R}}_{f}}}} \right),{{{{x}'_{i+1}}}}\left( {{c}_{{{\mathcal{R}}_{f}}}} \right) \right) \right)-1}}}+{{2}^{l\left( E\left( {{{{x}'_{m+1}}}}\left( {{c}_{{{\mathcal{R}}_{f}}}} \right),y\left( {{c}_{{{\mathcal{R}}_{f}}}} \right) \right) \right)-1}}, \\ 
\end{split}
\end{equation} 
where $l\left( E\left( x,y \right) \right)$ identifies the level of the entangled connection ${{E}_{{{\text{L}}_{l}}}}\left( x,y \right)$.

The probability of \eqref{ZEqnNum668817} is derived further as follows. At a given random $c_{{\rm {\mathcal R}}_{f} } $ and $r_{{\rm {\mathcal R}}_{f} } $, the probability that a given element (e.g., node or connection) $i$ is affected \cite{ref16} by the complex failure $f$ is defined as
\begin{equation} \label{ZEqnNum815504} 
\Pr (d_{i,c_{{\rm {\mathcal R}}_{f} } })=\left\{\begin{array}{l} {\frac{-d_{i,c_{{\rm {\mathcal R}}_{f} } } }{r_{{\rm {\mathcal R}}_{f} } } +1,\text{if}{\rm \; }d_{i,c_{{\rm {\mathcal R}}_{f} } } \le r_{{\rm {\mathcal R}}_{f} } } \\ {0,{\rm \; otherwise\; \; \; \; \; \; \; \; \; \; \; \; \; \; \; }} \end{array}\right. , 
\end{equation} 
where $d_{i,c_{{\rm {\mathcal R}}_{f} } } $ is the distance of element $i$ from the center $c_{{\rm {\mathcal R}}_{f} } $ of complex failure domain ${\rm {\mathcal R}}_{f} $.

\subsection{Entanglement Accessibility Ratio}

Let set ${\rm {\mathcal S}}^{{\rm *}} $ refer to those entangled connections of $N$ for which the condition $c$ (see \eqref{ZEqnNum801212}) holds after a complex failure $f$. Let $\Phi ^{c} \left(f\right)$ be a random variable that quantifies the ratio of total entanglement throughput in a complex failure event at a given $c$ (see \eqref{ZEqnNum801212}). This quantity is referred as the entanglement accessibility ratio (EAR) after a complex failure $f$ and identified by the ratio of total entanglement throughput after a complex failure $f$ of $N$ and the total entanglement throughput without a failure event \cite{ref16} at a given lower bound condition \eqref{ZEqnNum801212} as
\begin{equation} \label{ZEqnNum456523} 
\Phi ^{c} \left(f\right)=\frac{\sum _{i=1}^{\left|{\rm {\mathcal S}}^{{\rm *}} \right|}B_{F} (E_{{\rm L}_{l} }^{i}) }{\sum _{i=1}^{\left|{\rm {\mathcal S}}\right|}B_{F} (E_{{\rm L}_{l} }^{i}) } ,                                               
\end{equation} 
where $\left|{\rm {\mathcal S}}\right|$ is the number of connections in the set ${\rm {\mathcal S}}$ of $N$, and $\left|{\rm {\mathcal S}}^{{\rm *}} \right|$ is the cardinality of connection set ${\rm {\mathcal S}}^{{\rm *}} $ after a failure $f$ occurs in ${\rm {\mathcal R}}_{f} $.

\subsection{Cumulative Probability of Entanglement Accessibility Ratio}

Let $x$ be a critical lower bound on the entanglement accessibility ratio of $\Phi ^{c} \left(f\right)$ (see \eqref{ZEqnNum456523}) at a given condition $c$ and a complex failure $f$. A $\sigma \left(\Phi ^{c} \left(f\right)\right)$ cumulative probability of all complex failure events' occurrence for which the yielding ratio $\Phi ^{c} \left(f\right)$ at a given $c$ is at least $x$ (see \eqref{ZEqnNum456523}),
\begin{equation} \label{ZEqnNum652750} 
\Phi ^{c} \left(f\right)\ge x,                                                         
\end{equation} 
is referred to as the cumulative probability of entanglement accessibility ratio (CP-EAR) $\sigma ^{c} \left(\Phi ^{c} \left(f\right)\right)$, defined as
\begin{equation} \label{ZEqnNum490007} 
\begin{split}
   {{\sigma }^{c}}\left( {{\Phi }^{c}}\left( f \right) \right)&=\sum\limits_{f:{{\Phi }^{c}}\left( f \right)\ge x}{\Pr \left( f \right)} \\ 
 & =1-\sum\limits_{f:{{\Phi }^{c}}\left( f \right)<x}{\Pr \left( f \right)} \\ 
 & =1-{{\zeta }^{c}}\left( {{\Phi }^{c}}\left( f \right) \right),  
\end{split}
\end{equation} 
where $\zeta ^{c} \left(\Phi ^{c} \left(f\right)\right)$ is the cumulative distribution function of $\Phi ^{c} \left(f\right)$ at a condition $c$. 

The $\xi ^{c} \left(\Phi ^{c} \left(f\right)\right)$ probability density function (PDF) of ratio 
\begin{equation} \label{13)} 
\Phi ^{c} \left(f\right)=x 
\end{equation} 
after a complex failure $f$ is therefore
\begin{equation} \label{ZEqnNum572820} 
\xi ^{c} \left(\Phi ^{c} \left(f\right)\right)=\sum _{f:\Phi ^{c} \left(f\right)=x}\Pr \left(f\right) .                                            
\end{equation}

\subsection{Probabilistic Reduction of Entanglement Accessibility Ratio}

Assume that the $\zeta ^{c} \left(\Phi ^{c} \left(f\right)\right)$ cumulative distribution function of $\Phi ^{c} \left(f\right)$ at a condition $c$ is given as
\begin{equation} \label{ZEqnNum722837} 
\zeta ^{c} \left(\Phi ^{c} \left(f\right)\right)=\sum _{f:\Phi ^{c} \left(f\right)<x}\Pr \left(f\right) =q.                                        
\end{equation} 

Using \eqref{ZEqnNum722837}, the probabilistic reduction of entanglement accessibility ratio (PR-EAR) $\Omega ^{c} \left(\Phi ^{c} \left(f\right)\right)$ at a given ratio $x$, condition $c$, and probability $q$ is defined as
\begin{equation} \label{ZEqnNum498198} 
\begin{split}
   {{\Omega }^{c}}\left( {{\Phi }^{c}}\left( f \right) \right)&=\min \left\{ {{\Phi }^{c}}\left( f \right):{{\zeta }^{c}}\left( {{\Phi }^{c}}\left( f \right) \right)=q \right\} \\ 
 & =\min \left\{ {{\Phi }^{c}}\left( f \right):\sum\limits_{f:{{\Phi }^{c}}\left( f \right)<x}{\Pr \left( f \right)}=q \right\}.  
\end{split}
\end{equation} 

As follows, the PR-EAR parameter $\Omega ^{c} \left(x\right)$ in \eqref{ZEqnNum498198} quantifies the probability $q$ that the total entanglement accessibility ratio is reduced to at most ratio $x$ after a complex failure $f$.

\subsection{Domain-Dependent Entanglement Accessibility Ratio}

The $\Lambda ^{x} \left(r\right)$ domain-dependent entanglement accessibility ratio (DD-EAR) quantifies the $\Phi ^{c} \left(f\right)$ accessible entanglement ratio after a complex failure $f$ in a particular domain ${\rm {\mathcal R}}_{f} $ in a function of the radius $r_{{\rm {\mathcal R}}_{f} } $ of ${\rm {\mathcal R}}_{f} $ as
\begin{equation} \label{ZEqnNum448683} 
\Lambda ^{x} \left(r_{{\rm {\mathcal R}}_{f} } \right)=\sum _{\Phi ^{c} \left(f\right)}\Phi ^{c} \left(f\right)\varphi \left(\Phi ^{c} \left(f\right),r_{{\rm {\mathcal R}}_{f} } \right) ,                                     
\end{equation} 
where $\varphi \left(\Phi ^{c} \left(f\right),r_{{\rm {\mathcal R}}_{f} } \right)$ is the PDF of ratio $\Phi ^{c} \left(f\right)$ at an $r_{{\rm {\mathcal R}}_{f} } $-radius length complex failure domain ${\rm {\mathcal R}}_{f} $, defined as
\begin{equation} \label{18)} 
\varphi \left(\Phi ^{c} \left(f\right),r_{{\rm {\mathcal R}}_{f} } \right)=\sum _{f:\Phi ^{c} \left(f\right)=x,r_{{\rm {\mathcal R}}_{f} } }\Pr \left(f\right) .                                      
\end{equation} 
A complex network failure situation of a quantum repeater network $N$ with failure domain ${\rm {\mathcal R}}_{f} $ is illustrated in Fig. 1. A complex failure $f$ is associated with domain ${\rm {\mathcal R}}_{f} $, $f=1,\ldots ,m$. In the center $c_{{\rm {\mathcal R}}_{f} } $ of the ${\rm {\mathcal R}}_{f} $, for all $E_{{\rm L}_{l} }^{i} $ connections $B_{F} (E_{{\rm L}_{l} }^{i})=0$, and $\Pr \left(f\right)=1$. As the distance $d$ from the center of ${\rm {\mathcal R}}_{f} $ increases, the failure probability decreases, e.g., $\Pr \left(f\right)<1$. The condition $c:B_{F} (E_{{\rm L}_{l} }^{i})\ge B_{F}^{{\rm *}} (E_{{\rm L}_{l} }^{i})$ holds for $\forall i$, where $B_{F}^{{\rm *}} (E_{{\rm L}_{l} }^{i})$ is a critical lower bound on an $i$-th ${\rm L}_{l} $-level entangled connection $E_{{\rm L}_{l} }^{i} $, for the established entangled connections of $N$,

\begin{center}
\begin{figure*}[!h]
\begin{center}
\includegraphics[angle = 0,width=1\linewidth]{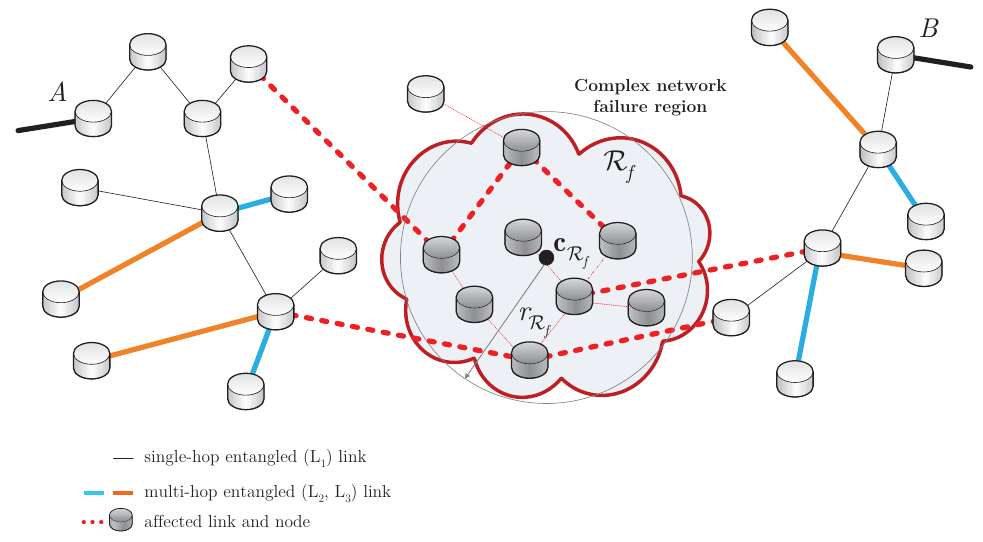}
\caption{An $f$ complex network failure scenario in a quantum Internet setting. A complex failure defines a domain ${\rm {\mathcal R}}_{f} $ (depicted by the gray-line circle) with a random radius $r_{{\rm {\mathcal R}}_{f} } $ and center $c_{{\rm {\mathcal R}}_{f} } $, and with a set of affected quantum nodes (depicted by dark gray nodes) and entangled connections (depicted by dashed red lines) between a source ($A$) quantum node and a target ($B$) quantum node (the affected network components are depicted by the gray cloud).} 
 \label{fig1}
 \end{center}
\end{figure*}
\end{center}

\section{Evaluation of Entanglement Accessibility}
\label{sec4}
In this section, first, we define a coefficient that describes the occurrence of a given entanglement accessibility ratio after a multiple complex failure scenario. Then we propose an empirical method to determine the occurrence coefficient from the observable quantities of a particular quantum network of the quantum Internet.

\subsection{Occurrence Coefficient}

Let $Q\left(\Phi ^{c} \left(f\right)\right)$ refer to the occurrence coefficient of a particular $\Phi ^{c} \left(f\right)$ entanglement accessibility ratio at a complex failure domain ${\rm {\mathcal R}}_{f} $ in $N$, defined as
\begin{equation} \label{ZEqnNum101546} 
Q\left(\Phi ^{c} \left(f\right)\right)=\frac{{\rm {\mathcal N}}\left(\Phi ^{c} \left(f\right)\right)}{{\rm {\mathcal N}}\left({\rm {\mathcal A}}^{c} \left(f\right)\right)} , 
\end{equation} 
where ${\rm {\mathcal N}}\left(\Phi ^{c} \left(f\right)\right)$ is the number of occurrence of a given entanglement accessibility ratio $\Phi ^{c} \left(f\right)$ in $N$ after a failure $f$, while ${\rm {\mathcal N}}\left({\rm {\mathcal A}}^{c} \left(f\right)\right)$ quantifies the total number of occurrences of all accessible ratios ${\rm {\mathcal A}}^{c} \left(f\right)$ in $N$ after a failure $f$. 

Extending \eqref{ZEqnNum101546} to all the $m$ complex failure domains ${\rm {\mathcal R}}_{f=1} ,\ldots ,{\rm {\mathcal R}}_{f=m} $ yields 
\begin{equation} \label{ZEqnNum752231} 
\begin{split}
   {{Q}^{tot}}(N)&=\sum\limits_{f}{Q\left( {{\Phi }^{c}}\left( f \right) \right)} \\ 
 & ={{Q}^{\left( f=1 \right)}}\left( {{\Phi }^{c}}\left( f \right) \right)+\ldots +{{Q}^{\left( f=m \right)}}\left( {{\Phi }^{c}}\left( f \right) \right),  
\end{split}
\end{equation} 
where $Q^{\left(f=i\right)} \left(\Phi ^{c} \left(f\right)\right)$ quantifies the occurrence of ratio $\Phi ^{c} \left(f\right)$ via \eqref{ZEqnNum101546} for an $i$-th domain ${\rm {\mathcal R}}_{f=i} $. 

In the function of $Q^{tot}(N)$, the quantities of \eqref{ZEqnNum490007}, \eqref{ZEqnNum498198}, and \eqref{ZEqnNum448683} can be derived as follows. 

For an $m$-domain setting with domains ${\rm {\mathcal R}}_{f=1} ,\ldots ,{\rm {\mathcal R}}_{f=m} $, $\sigma ^{c} \left(\Phi ^{c} \left(f\right)\right)$ can be derived from the function $Q^{tot}(N)$ as
\begin{equation} \label{ZEqnNum560653} 
\sigma ^{c} \left(\Phi ^{c} \left(f\right)\right)=\frac{Q^{tot}(N)}{m} , 
\end{equation} 
while $\Omega ^{c} \left(\Phi ^{c} \left(f\right)\right)$ at ${\rm {\mathcal R}}_{f=1} ,\ldots ,{\rm {\mathcal R}}_{f=m} $ is
\begin{equation} \label{22)} 
\Omega ^{c} \left(\Phi ^{c} \left(f\right)\right)=\min \left\{\Phi ^{c} \left(f\right):\frac{1-Q^{tot}(N)}{m} =q\right\}. 
\end{equation} 

At a particular failure domain radius $r_{{\rm {\mathcal R}}_{f} } $ of a given ${\rm {\mathcal R}}_{f} $, let 
\begin{equation} \label{ZEqnNum471721} 
\begin{split}
   \tilde{Q}\left( {{\Phi }^{c}}\left( f \right),{{r}_{{{\mathcal{R}}_{f}}}} \right)&={{\xi }^{c}}\left( {{\Phi }^{c}}\left( f \right),{{r}_{{{\mathcal{R}}_{f}}}} \right) \\ 
 & =\sum\limits_{f:{{\Phi }^{c}}\left( f \right)=x,{{r}_{{{\mathcal{R}}_{f}}}}}{\Pr \left( f \right),}  
\end{split}
\end{equation} 
where $\xi ^{c} \left(\Phi ^{c} \left(f\right)\right)$ as shown in \eqref{ZEqnNum572820}. 

For all domains ${\rm {\mathcal R}}_{f=1} ,\ldots ,{\rm {\mathcal R}}_{f=m} $, \eqref{ZEqnNum471721} extends to
\begin{equation} \label{24)} 
\begin{split}
   {{{\tilde{Q}}}^{tot}}\left( {{\Phi }^{c}}\left( f \right),{{r}_{{{\mathcal{R}}_{f}}}} \right)&=\sum\limits_{f}{\tilde{Q}\left( {{\Phi }^{c}}\left( f \right),{{r}_{{{\mathcal{R}}_{f}}}} \right)} \\ 
 & ={{{\tilde{Q}}}^{\left( f=1 \right)}}\left( {{\Phi }^{c}}\left( f \right),{{r}_{{{\mathcal{R}}_{f}}}} \right)+\ldots +{{{\tilde{Q}}}^{\left( f=1 \right)}}\left( {{\Phi }^{c}}\left( f \right),{{r}_{{{\mathcal{R}}_{f}}}} \right),  
\end{split}
\end{equation} 
where $\tilde{Q}^{\left(f=i\right)} \left(\Phi ^{c} \left(f\right),r_{{\rm {\mathcal R}}_{f} } \right)$ quantifies the occurrence of ratio $\Phi ^{c} \left(f\right)$ for an $i$-th domain ${\rm {\mathcal R}}_{f=i} $ via a particular radius $r_{{\rm {\mathcal R}}_{f} } $ using \eqref{ZEqnNum471721}. Then $\Lambda ^{x} \left(r_{{\rm {\mathcal R}}_{f} } \right)$ in a ${\rm {\mathcal R}}_{f=1} ,\ldots ,{\rm {\mathcal R}}_{f=m} $ scenario is expressed as
\begin{equation} \label{ZEqnNum754105} 
\Lambda ^{x} \left(r_{{\rm {\mathcal R}}_{f} } \right)=\sum _{\Phi ^{c} \left(f\right)}\Phi ^{c} \left(f\right)\frac{\tilde{Q}^{tot} \left(\Phi ^{c} \left(f\right),r_{{\rm {\mathcal R}}_{f} } \right)}{m}  .                                 
\end{equation} 

Therefore, \eqref{ZEqnNum560653} to \eqref{ZEqnNum754105} follow that the entanglement accessibility ratios can be determined via the occurrence coefficient $\tilde{Q}^{tot} \left(\Phi ^{c} \left(f\right),r_{{\rm {\mathcal R}}_{f} } \right)$. 

To find this quantity at a given network $N$ empirically, we propose an algorithm as follows.

\subsection{Empirical Evaluation of Occurrence Coefficient}

We propose an algorithm, ${\rm {\mathcal A}}_{Q\left(\Phi ^{c} \left(f\right)\right)} $, for the empirical determination of the O-EAR coefficient $Q\left(\Phi ^{c} \left(f\right)\right)$ (see \eqref{ZEqnNum101546}) at a complex failure domain ${\rm {\mathcal R}}_{f} $ scenario and then the evaluation of $Q^{tot}(N)$ (see \eqref{ZEqnNum752231}) by the extended analysis of all domains ${\rm {\mathcal R}}_{f=1} ,\ldots ,{\rm {\mathcal R}}_{f=m} $. Some preliminary definitions are as follows.

\subsubsection{Definitions}

To describe the topology of $N$, let $I_{N} $ be the node-to-node incidence matrix of $N$, and let $\tilde{I}_{N} $ refer to a temporal incidence matrix for the iteration steps of the algorithm.

Each $L_{i} $-level entangled connection is characterized by a particular entanglement throughput rate $B_{F} (E_{{\rm L}_{l} }^{i})$, which are used to determine the $A\left({\rm {\mathcal S}}\right)$ total accessible entanglement at a connection set ${\rm {\mathcal S}}$ at no failure as 
\begin{equation} \label{ZEqnNum663073} 
A\left({\rm {\mathcal S}}\right)=\sum _{i=1}^{\left|{\rm {\mathcal S}}\right|}B_{F} (E_{{\rm L}_{l} }^{i}) .                                               
\end{equation} 

Then let $A_{\rho ,U_{k} } $ and $B_{\rho ,U_{k} } $ be the source and target quantum nodes of a demand $\rho $ associated to user $U_{k} $, $k=1,\ldots ,K$, where $K$ is the number of users. Then let $D\left(\rho \left({\rm {\mathcal S}}'\right)\right)$ be the total required entanglement by a demand $\rho $ as 
\begin{equation} \label{ZEqnNum643627} 
D\left(\rho \left({\rm {\mathcal S}}'\right)\right)=\sum _{i=1}^{\left|\rho \left({\rm {\mathcal S}}'\right)\right|}B_{F} (E_{{\rm L}_{l} }^{i}) ,                                           
\end{equation} 
f $\rho \left({\rm {\mathcal S}}'\right)$ refers to the connection set ${\rm {\mathcal S}}'$ of $\rho $. 

For a given demand $\rho _{i} $, let 
\begin{equation} \label{ZEqnNum216010} 
D^{{\rm {\mathcal P}}\left(N\right)} \left(\rho _{i} \left({\rm {\mathcal S}}'_{i} \right)\right) 
\end{equation} 
quantify the total required entanglement of demand $\rho _{i} $ with connection set ${\rm {\mathcal S}}'_{i} $ along entangled connections traversed by respective paths ${\rm {\mathcal P}}\left(N\right)$ in $N$. 

Let 
\begin{equation} \label{ZEqnNum619590} 
\mho =\left\{\rho _{1} ,\ldots , \rho _{g} \right\} 
\end{equation} 
identify a set of $g$ demands with both end nodes $A_{\rho \in \mho , U_{k} } $ and $B_{\rho \in \mho ,U_{k} } $ not affected by a complex failure $f$. 

Assuming that a complex failure $f$ with a domain ${\rm {\mathcal R}}_{f} $ occurs in $N$, the total accessible entanglement after a complex failure $f$ is
\begin{equation} \label{ZEqnNum538510} 
A\left({\rm {\mathcal S}}^{{\rm *}} \right)=\sum _{i=1}^{\left|{\rm {\mathcal S}}^{{\rm *}} \right|}B_{F} (E_{{\rm L}_{l} }^{i}) ,                                          
\end{equation} 
where ${\rm {\mathcal S}}^{{\rm *}} $ is the connection set of $N$ after the failure. 

The center $c_{{\rm {\mathcal R}}_{f} } $ of a domain ${\rm {\mathcal R}}_{f} $ and the corresponding radius length $r_{{\rm {\mathcal R}}_{f} } $ of ${\rm {\mathcal R}}_{f} $ are modeled as uniformly distributed random continuous variables \cite{ref16}. 

At a given $\hat{B}_{F} (E_{{\rm L}_{l} }^{i})$ upper bound on the entanglement throughput of $E_{{\rm L}_{l} }^{i} $, the remaining accessible entanglement throughput is defined as
\begin{equation} \label{ZEqnNum762617} 
F(E_{{\rm L}_{l} }^{i})=\hat{B}_{F} (E_{{\rm L}_{l} }^{i})-B_{F} (E_{{\rm L}_{l} }^{i}),                                      
\end{equation} 
where $B_{F} (E_{{\rm L}_{l} }^{i})$ refers to a current rate.

Let $R_{f} \left(N\right)$ quantify the empirical estimate of entanglement accessible ratio $\Phi ^{c} \left(f\right)$ (see \eqref{ZEqnNum456523}) after a complex failure $f$ in a given quantum network $N$, as 
\begin{equation} \label{ZEqnNum624971} 
{{R}_{f}}\left( N \right)=\frac{A\left( {{\mathcal{S}}^{*}} \right)}{A\left( \mathcal{S} \right)},
\end{equation} 
where $A\left({\rm {\mathcal S}}^{{\rm *}} \right)$ is defined in \eqref{ZEqnNum538510}, while $A\left({\rm {\mathcal S}}\right)$ is given by \eqref{ZEqnNum663073}. Therefore, $R_{f} \left(N\right)$ provides an estimation of $Q\left(\Phi ^{c} \left(f\right)\right)$ from the empirical values of \eqref{ZEqnNum538510} and \eqref{ZEqnNum663073} as
\begin{equation} \label{uj33)} 
Q\left(\Phi ^{c} \left(f\right)\right)=\frac{{\rm {\mathcal N}}(R_{f} \left(N\right))}{{\rm {\mathcal N}}\left({\rm {\mathcal A}}^{c} \left(f\right)\right)} , 
\end{equation}

\subsubsection{Algorithm}

The ${\rm {\mathcal A}}_{Q\left(\Phi ^{c} \left(f\right)\right)} $ algorithm aims to determine the empirical estimation of the occurrence function $Q\left(\Phi ^{c} \left(f\right)\right)$. 

The algorithm ${\rm {\mathcal A}}_{Q\left(\Phi ^{c} \left(f\right)\right)} $ for a ${\rm {\mathcal R}}_{f=1} ,\ldots ,{\rm {\mathcal R}}_{f=m} $ multiple complex failure scenario is given in Algorithm 1.

  \setcounter{algocf}{0}
\begin{algo}
  \DontPrintSemicolon
\caption{Estimation of Occurrence of Entanglement Accessibility Ratio}

\textbf{Step 1}. Let $\tilde{I}_{N} =I_{N} $ and $A\left({\rm {\mathcal S}}^{{\rm *}} \right)=0$. At a given $f$, determine $\Pr (d_{i,c_{{\rm {\mathcal R}}_{f} } })$ for all $i$. For all connections of ${\rm {\mathcal S}}$ for which condition $c$ does not hold, set the corresponding elements of $\tilde{I}_{N} $ to 0. 

\textbf{Step 2}. For all entangled connections of ${\rm {\mathcal S}}^{{\rm *}} $, set $F(E_{{\rm L}_{l} }^{i})=\hat{B}_{F} (E_{{\rm L}_{l} }^{i})$. For all $\rho _{i} $ demands of $\mho $, set 
\[D^{{\rm {\mathcal P}}\left(N\right)} \left(\rho _{i} \left({\rm {\mathcal S}}'_{i} \right)\right)=D\left(\rho _{i} \left({\rm {\mathcal S}}'_{i} \right)\right).\]
Using $\tilde{I}_{N} $, determine the shortest path $\dot{{\rm {\mathcal P}}}_{i} $ for demand $\rho _{i} $.

\textbf{Step 3}. For all $\rho _{i} $ of $\mho $, evaluate 
\[A\left( {{{{\mathcal{S}}'_{i}}}} \right)=\underset{{{{\dot{\mathcal{P}}}}_{i}}\in E_{{{\text{L}}_{l}}}^{i}}{\mathop{\min }}\,F\left( E_{{{\text{L}}_{l}}}^{i} \right)=\underset{{{{\dot{\mathcal{P}}}}_{i}}\in E_{{{\text{L}}_{l}}}^{i}}{\mathop{\min }}\,{{\hat{B}}_{F}}\left( E_{{{\text{L}}_{l}}}^{i} \right)\]
If $D^{{\rm {\mathcal P}}\left(N\right)} \left(\rho _{i} \left({\rm {\mathcal S}}'_{i} \right)\right)\le A\left({\rm {\mathcal S}}'_{i} \right)$, then set 
\[A\left({\rm {\mathcal S}}^{{\rm *}} \right)=A\left({\rm {\mathcal S}}^{{\rm *}} \right)+D^{{\rm {\mathcal P}}\left(N\right)} \left(\rho _{i} \left({\rm {\mathcal S}}'_{i} \right)\right),\] 
and set 
\[D^{{\rm {\mathcal P}}\left(N\right)} \left(\rho _{i} \left({\rm {\mathcal S}}'_{i} \right)\right)=0.\]

For all entangled connections traversed by $\dot{{\rm {\mathcal P}}}_{i} $, set $F(E_{{\rm L}_{l} }^{i})=F(E_{{\rm L}_{l} }^{i})-D^{{\rm {\mathcal P}}\left(N\right)} \left(\rho _{i} \left({\rm {\mathcal S}}'_{i} \right)\right)$. 
If $D^{{\rm {\mathcal P}}\left(N\right)} \left(\rho _{i} \left({\rm {\mathcal S}}'_{i} \right)\right)>A\left({\rm {\mathcal S}}'_{i} \right)$, then set 
\[A\left({\rm {\mathcal S}}^{{\rm *}} \right)=A\left({\rm {\mathcal S}}^{{\rm *}} \right)+A\left({\rm {\mathcal S}}'_{i} \right),\] 
and set 
\[D^{{\rm {\mathcal P}}\left(N\right)} \left(\rho _{i} \left({\rm {\mathcal S}}'_{i} \right)\right)=D^{{\rm {\mathcal P}}\left(N\right)} \left(\rho _{i} \left({\rm {\mathcal S}}'_{i} \right)\right)-A\left({\rm {\mathcal S}}'_{i} \right).\]  
For all entangled connections traversed by $\dot{{\rm {\mathcal P}}}_{i} $ set 
\[F(E_{{\rm L}_{l} }^{i})=F(E_{{\rm L}_{l} }^{i})-A\left({\rm {\mathcal S}}'_{i} \right).\]

\textbf{Step 4}. Define a set of demands $\lambda $, which contains all $\rho _{i} $ demands, where $D^{{\rm {\mathcal P}}\left(N\right)} \left(\rho _{i} \left({\rm {\mathcal S}}'_{i} \right)\right)>0$. Determine the next shortest path $\ddot{{\rm {\mathcal P}}}_{i} $. Set $A\left({\rm {\mathcal S}}^{{\rm *}} \right)=A\left({\rm {\mathcal S}}^{{\rm *}} \right)+X$ and $D^{{\rm {\mathcal P}}\left(N\right)} \left(\rho _{i} \left({\rm {\mathcal S}}'_{i} \right)\right)=D^{{\rm {\mathcal P}}\left(N\right)} \left(\rho _{i} \left({\rm {\mathcal S}}'_{i} \right)\right)-X$, where $X$ is a given ratio of the maximum of the total accessible entanglement throughput of the entangled connections of $\ddot{{\rm {\mathcal P}}}_{i} $. For all $E_{{\rm L}_{l} }^{i} $ entangled connections traversed by $\ddot{{\rm {\mathcal P}}}_{i} $, determine the current $F(E_{{\rm L}_{l} }^{i})$. 

\textbf{Step 5}. Repeat step 4 until $D^{{\rm {\mathcal P}}\left(N\right)} \left(\rho _{i} \left({\rm {\mathcal S}}'_{i} \right)\right)=0$ or $\ddot{{\rm {\mathcal P}}}_{i} =\emptyset $ holds. Output $R_{f} \left(N\right)$ via \eqref{ZEqnNum624971}, and the empirical estimation of the $Q\left(R_{f} \left(N\right)\right)$ occurrence from \eqref{ZEqnNum101546} via \eqref{uj33)}.

\textbf{Step 6}. Repeat steps 1 to 5 for all $m$ complex failure domains ${\rm {\mathcal R}}_{f=1} ,\ldots ,{\rm {\mathcal R}}_{f=m} $ and output $Q^{tot}(N)$ via \eqref{ZEqnNum752231}.

\end{algo}

\subsection{Description}

A brief description of the ${\rm {\mathcal A}}_{Q\left(\Phi ^{c} \left(f\right)\right)} $ method is as follows. In steps 1 and 2, some initializations are performed for further calculations. Steps 3 to 5 derive the ratio $R_{f} \left(N\right)\approx \Phi ^{c} \left(f\right)$ of accessible entanglement at a given failure domain ${\rm {\mathcal R}}_{f} $. The steps aim to determine the ratio of total accessible entanglement in a given complex domain failure scenario. For each demand that has unaffected end nodes, a path searching is performed to find the shortest alternate path $\dot{{\rm {\mathcal P}}}_{i} $ for all demands $\rho _{i} $ to serve requirement $D\left(\rho _{i} \left({\rm {\mathcal S}}'_{i} \right)\right)$ of a given $\rho _{i} $. If an alternate path exists but the entangled connections of the path are not able to serve the required entanglement $D\left(\rho _{i} \left({\rm {\mathcal S}}'_{i} \right)\right)$, then a new shortest path $\ddot{{\rm {\mathcal P}}}_{i} $ is determined. The calculations are performed for all demands that are present with a nonzero required entanglement in the network. In step 6, the iteration is extended for the evaluation of all failure domains ${\rm {\mathcal R}}_{f=1} ,\ldots ,{\rm {\mathcal R}}_{f=m} $.

\subsubsection{Step 1}

In step 1, a temporal incidence matrix $\tilde{I}_{N} $ is initialized by $I_{N} $, and the value of the total accessible entanglement via set ${\rm {\mathcal S}}^{{\rm *}} $ after a complex failure $f$ is set to zero, $A\left({\rm {\mathcal S}}^{{\rm *}} \right)=0$, where $A\left({\rm {\mathcal S}}^{{\rm *}} \right)$ is defined in \eqref{ZEqnNum538510}. To identify the set of quantum nodes affected by $f$, for all nodes their corresponding probability $\Pr (d_{i,c_{{\rm {\mathcal R}}_{f} } })$ is determined via \eqref{ZEqnNum815504} in a function of distance $d_{i,c_{{\rm {\mathcal R}}_{f} } } $ node $i$ from center $c_{{\rm {\mathcal R}}_{f} } $ of ${\rm {\mathcal R}}_{f} $. Then to distinguish the unusable connections after $f$ has occurred for all connections for which condition $c$ does not hold (see \eqref{ZEqnNum801212}), set the corresponding elements of $\tilde{I}_{N} $ to 0.

\subsubsection{Step 2}

In step 2, for all entangled connections of ${\rm {\mathcal S}}^{{\rm *}} $, the amount of the utilizable throughput rate is set to a maximum of the given entangled connection $E_{{\rm L}_{l} }^{i} $, $F(E_{{\rm L}_{l} }^{i})=\hat{B}_{F} (E_{{\rm L}_{l} }^{i})$, where $\hat{B}_{F} (E_{{\rm L}_{l} }^{i})$, the upper bound on the throughput of an entangled connection $E_{{\rm L}_{l} }^{i} $, and $F(E_{{\rm L}_{l} }^{i})$ are given by \eqref{ZEqnNum762617}. Initialize a set $\mho =\left\{\rho _{1} ,\ldots ,\rho _{g} \right\}$ of demands with both end nodes $A_{\rho \in \mho ,U_{k} } $ and $B_{\rho \in \mho ,U_{k} } $ not affected by $f$ as given by \eqref{ZEqnNum619590}. The quantity of $D^{{\rm {\mathcal P}}\left(N\right)} \left(\rho _{i} \left({\rm {\mathcal S}}'_{i} \right)\right)$ (see \eqref{ZEqnNum216010}), which describes the required total entanglement by demand $\rho _{i} $ with connection set ${\rm {\mathcal S}}'_{i} $ along entangled connections traversed by respective paths ${\rm {\mathcal P}}\left(N\right)$ in $N$, is set to the amount of the total entanglement required for $\rho _{i} $, $D\left(\rho _{i} \left({\rm {\mathcal S}}'_{i} \right)\right)$ (see \eqref{ZEqnNum643627}). As a final substep, determine the shortest path $\dot{{\rm {\mathcal P}}}_{i} $ for $\rho _{i} $ by using the temporarily incidence matrix $\tilde{I}_{N} $ as characterized in step 1.

\subsubsection{Step 3}

In step 3, some computations are performed for the demands $\rho _{i} $ of set $\mho $, whose demands are not affected by the failure. The value of the total accessible entanglement via connection set ${\rm {\mathcal S}}'_{i} $ of a given demand $\rho _{i} $ after a complex failure $f$, $A\left({\rm {\mathcal S}}'_{i} \right)$ (see \eqref{ZEqnNum663073}), is set to the minimal amount of utilizable throughput rate of $\dot{{\rm {\mathcal P}}}_{i} $, thus 
\begin{equation}
A\left({\rm {\mathcal S}}'_{i} \right)=\mathop{\min }\limits_{\dot{{\rm {\mathcal P}}}_{i} \in E_{{\rm L}_{l} }^{i} } F(E_{{\rm L}_{l} }^{i}). 
\end{equation}
From step 2, it follows that $A\left({\rm {\mathcal S}}'_{i} \right)$ will be equal to the maximal entanglement rate of that entangled connection, which yields the min-max optimization 
\begin{equation}
A\left({\rm {\mathcal S}}'_{i} \right)=\mathop{\min }\limits_{\dot{{\rm {\mathcal P}}}_{i} \in E_{{\rm L}_{l} }^{i} } \hat{B}_{F} (E_{{\rm L}_{l} }^{i}). 
\end{equation}
After this substep, the relation of $D^{{\rm {\mathcal P}}\left(N\right)} \left(\rho _{i} \left({\rm {\mathcal S}}'_{i} \right)\right)$ and $A\left({\rm {\mathcal S}}'_{i} \right)$ is verified, and the next steps are selected based on it. If the value of the required total entanglement $D^{{\rm {\mathcal P}}\left(N\right)} \left(\rho _{i} \left({\rm {\mathcal S}}'_{i} \right)\right)$ of demand $\rho _{i} $ along entangled connections traversed by respective paths ${\rm {\mathcal P}}\left(N\right)$ in $N$ does not exceed $A\left({\rm {\mathcal S}}'_{i} \right)$, the value of the total accessible entanglement of demand $\rho _{i} $ after a complex failure $f$, then $A\left({\rm {\mathcal S}}^{{\rm *}} \right)$ value of total accessible entanglement via connection set ${\rm {\mathcal S}}^{{\rm *}} $ after a complex failure $f$ is increased by $D^{{\rm {\mathcal P}}\left(N\right)} \left(\rho _{i} \left({\rm {\mathcal S}}'_{i} \right)\right)$. Conversely, if $D^{{\rm {\mathcal P}}\left(N\right)} \left(\rho _{i} \left({\rm {\mathcal S}}'_{i} \right)\right)$ exceeds $A\left({\rm {\mathcal S}}'_{i} \right)$, then $A\left({\rm {\mathcal S}}^{{\rm *}} \right)$ is increased by $A\left({\rm {\mathcal S}}'_{i} \right)$. As the value of $A\left({\rm {\mathcal S}}^{{\rm *}} \right)$ is determined, depending on the relation of $D^{{\rm {\mathcal P}}\left(N\right)} \left(\rho _{i} \left({\rm {\mathcal S}}'_{i} \right)\right)$ and $A\left({\rm {\mathcal S}}'_{i} \right)$, the value of the required total entanglement $D^{{\rm {\mathcal P}}\left(N\right)} \left(\rho _{i} \left({\rm {\mathcal S}}'_{i} \right)\right)$ is either decreased by $D^{{\rm {\mathcal P}}\left(N\right)} \left(\rho _{i} \left({\rm {\mathcal S}}'_{i} \right)\right)$ or by $A\left({\rm {\mathcal S}}'_{i} \right)$. This substep therefore yields $D^{{\rm {\mathcal P}}\left(N\right)} \left(\rho _{i} \left({\rm {\mathcal S}}'_{i} \right)\right)=0$ if $D^{{\rm {\mathcal P}}\left(N\right)} \left(\rho _{i} \left({\rm {\mathcal S}}'_{i} \right)\right)\le A\left({\rm {\mathcal S}}'_{i} \right)$, but results in $D^{{\rm {\mathcal P}}\left(N\right)} \left(\rho _{i} \left({\rm {\mathcal S}}'_{i} \right)\right)=D^{{\rm {\mathcal P}}\left(N\right)} \left(\rho _{i} \left({\rm {\mathcal S}}'_{i} \right)\right)-A\left({\rm {\mathcal S}}'_{i} \right)$ if $D^{{\rm {\mathcal P}}\left(N\right)} \left(\rho _{i} \left({\rm {\mathcal S}}'_{i} \right)\right)>A\left({\rm {\mathcal S}}'_{i} \right)$. Depending on the relation of $D^{{\rm {\mathcal P}}\left(N\right)} \left(\rho _{i} \left({\rm {\mathcal S}}'_{i} \right)\right)$ and $A\left({\rm {\mathcal S}}'_{i} \right)$, a final computation is also performed in this step. For each entangled connection traversed by the shortest path $\dot{{\rm {\mathcal P}}}_{i} $, the amount of remaining utilizable entanglement throughput is decreased as $F(E_{{\rm L}_{l} }^{i})=F(E_{{\rm L}_{l} }^{i})-D^{{\rm {\mathcal P}}\left(N\right)} \left(\rho _{i} \left({\rm {\mathcal S}}'_{i} \right)\right)$ if $D^{{\rm {\mathcal P}}\left(N\right)} \left(\rho _{i} \left({\rm {\mathcal S}}'_{i} \right)\right)\le A\left({\rm {\mathcal S}}'_{i} \right)$, and $F(E_{{\rm L}_{l} }^{i})=F(E_{{\rm L}_{l} }^{i})-A\left({\rm {\mathcal S}}'_{i} \right)$ if $D^{{\rm {\mathcal P}}\left(N\right)} \left(\rho _{i} \left({\rm {\mathcal S}}'_{i} \right)\right)>A\left({\rm {\mathcal S}}'_{i} \right)$ holds.

\subsubsection{Step 4}

In step 4, a set $\lambda $ of demands is determined via condition $D^{{\rm {\mathcal P}}\left(N\right)} \left(\rho _{i} \left({\rm {\mathcal S}}'_{i} \right)\right)>0$. It follows that some demanded entanglement cannot be served fully; thus, in this step, the entanglement assigned to the demands should be increased as much as possible. These demands are still associated with a nonzero required entanglement ratio in the network, and therefore, these queries should be processed. This step focuses on the service of these demands via the corresponding calculations that are similar to the calculations of step 3. The $A\left({\rm {\mathcal S}}^{{\rm *}} \right)$ value is increased by a given $X$, which is a given ratio of the maximum of the total accessible entanglement throughput of the entangled connections of the next shortest path $\ddot{{\rm {\mathcal P}}}_{i} $. Then the value of $D^{{\rm {\mathcal P}}\left(N\right)} \left(\rho _{i} \left({\rm {\mathcal S}}'_{i} \right)\right)$ is decreased by ratio $X$.

\subsubsection{Step 5}

In step 5, all demands are served until there is no nonzero required entanglement present in the network. All demands are served if $D^{{\rm {\mathcal P}}\left(N\right)} \left(\rho _{i} \left({\rm {\mathcal S}}'_{i} \right)\right)=0$ for all $\rho _{i} $. The serving process of demands also stops if there is no next shortest path $\ddot{{\rm {\mathcal P}}}_{i} $ in the network; therefore, $\ddot{{\rm {\mathcal P}}}_{i} =\emptyset $ holds. Finally, the empirical estimation of the ratio of accessible entanglement after a failure is determined as $R_{f} \left(N\right)={A\left({\rm {\mathcal S}}^{{\rm *}} \right) \mathord{\left/{\vphantom{A\left({\rm {\mathcal S}}^{{\rm *}} \right) A\left({\rm {\mathcal S}}\right)}}\right.\kern-\nulldelimiterspace} A\left({\rm {\mathcal S}}\right)} $ (see \eqref{ZEqnNum624971}). The estimation of  $Q\left(R_{f} \left(N\right)\right)$ (see \eqref{ZEqnNum101546}) uses the empirical value of $A\left({\rm {\mathcal S}}^{{\rm *}} \right)$ after a complex failure $f$ via connection set ${\rm {\mathcal S}}^{{\rm *}} $, and also the empirical value of the $A\left({\rm {\mathcal S}}\right)$ via connection set ${\rm {\mathcal S}}$. Using the resulting estimate $R_{f} \left(N\right)$ in \eqref{ZEqnNum624971}, $Q\left(R_{f} \left(N\right)\right)$ can be determined via the estimation in \eqref{uj33)}.

\subsubsection{Step 6}

Finally, step 6 extends the results for all the $m$ failure events occurring in $N$ to determine $Q^{tot}(N)$ (see \eqref{ZEqnNum752231}).

\subsection{Computational Complexity}

The computational complexity of algorithm ${\rm {\mathcal A}}_{Q\left(\Phi ^{c} \left(f\right)\right)} $ depends on the complexity of the searching method applied in steps 3 and 4 to compute the shortest paths. Using a base-graph method \cite{ref8,ref9,ref10} to determine the shortest path with respect to the entanglement throughput metric, the complexity of the method is at most ${\rm {\mathcal O}}\left(\log n\right)^{2} $, where $n$ is the size of a $k$-dimensional $n$-size base-graph $G^{k} $ of $N$.

\subsection{Non-Linear Optimization for the Control Observable}
A non-stochastic regulation (NSR) \cite{nsrq,nsr1,nsr2} non-linear optimization method can be defined within the proposed scheme to yield an estimation of the occurrence coefficient (control observable), in the following manner.

Let $Q\left(\Phi ^{c} \left(f\right)\right)$ be an actual occurrence ratio at a particular $f$ in $N$ subject to be estimated, and let 
\begin{equation} \label{1)} 
\vec{R} \left(N\right)=\left(R_{f=1} \left(N\right),\ldots ,R_{f=m} \left(N\right)\right)^{T}  
\end{equation} 
be the noisy empirical vector of the $R_{f} \left(N\right)$, $f=1,\ldots ,m$ noisy quantities associated with the $m$ failure domains ${\rm {\mathcal R}}_{1} ,\ldots ,{\rm {\mathcal R}}_{m} $. 

In the optimization model it is assumed that the empirical statistical information obtainable from the quantum network is noisy. Let $\Delta $ be a noise vector associated to the estimation error, such that
\begin{equation} \label{ZEqnNum227438} 
\vec{R} \left(N\right)=\vec{Q}^{tot} \left(N\right)+\Delta ,   
\end{equation} 
where $\vec{Q}^{tot} \left(N\right)$ is the vector as
\begin{equation} \label{3)} 
\vec{Q}^{tot} \left(N\right)=\left(Q\left(\Phi ^{c} \left(f=1\right)\right),\ldots ,Q\left(\Phi ^{c} \left(f=m\right)\right)\right)^{T} .   
\end{equation} 
Then, the $\left\langle Q^{tot} \left(N\right)\right\rangle $ estimate of $Q^{tot} \left(N\right)$ yielded via an NSR optimization \cite{nsrq,nsr1,nsr2} is as
\begin{equation} \label{ZEqnNum378013} 
\begin{split}
   \left\langle {{{\vec{Q}}}^{tot}}\left( N \right) \right\rangle =\arg \underset{Q\left( {{\Phi }^{c}}\left( f \right) \right)}{\mathop{\min }}\,&{{\left( \left( {{{\vec{R}}}}\left( N \right)-\xi \left( \vec{d}\otimes {{e}^{{{{\vec{Q}}}^{tot}}\left( N \right)}} \right) \right) \right.}^{T}}{{\left( {{K}_{\Delta }} \right)}^{-1}} \\ 
 & \times \left( {{{\vec{R}}}}\left( N \right)-\xi \left( \vec{d}\otimes {{e}^{{{{\vec{Q}}}^{tot}}\left( N \right)}} \right) \right) \\ 
 & +{{\omega }^{-2}}\left. \int\limits {{{\vec{Q}'^{tot} \left(N\right)}^{2}}dt} \right),  
\end{split}
\end{equation} 
where $\omega $ is an unknown regularization parameter, $\xi $ is a linear operator, $\vec{d}$ is a matrix, as
\begin{equation} \label{5)} 
\vec{d}=\left(d\left(f=1\right),\ldots ,d\left(f=m\right)\right)^{T} ,   
\end{equation} 
where $d\left(f\right)$ is a deterministic exponential function 
\begin{equation} \label{ZEqnNum652581} 
d\left(f\right)=\frac{1}{\delta } e^{\frac{-f}{\delta } },  
\end{equation} 
where $\delta $ is an unknown regularization parameter, such that from \eqref{ZEqnNum652581}
\begin{equation} \label{7)} 
\vec{Q}^{tot} \left(N\right)=d\left(f\right)\otimes e^{Q\left(\Phi ^{c} \left(f\right)\right)},  
\end{equation} 
where $Q\left(\Phi ^{c} \left(f\right)\right)$ is as
\begin{equation} \label{ZEqnNum788492} 
Q\left(\Phi ^{c} \left(f\right)\right)=\alpha +\gamma \varphi \left(\Delta \right),  
\end{equation} 
where $\alpha $ and $\gamma $ are unknown regularization parameters, $\varphi \left(\Delta \right)$ is a process that represents the noise of the empirical estimation; $K_{\Delta } $ is the covariance matrix of the noise $\Delta $ included in the empirical vector $\vec{R} \left(N\right)$, $\gamma $ is a regularization parameter, $\vec{Q}'^{tot} \left(N\right)$ is the derivative of $\vec{Q}^{tot} \left(N\right)$, while $\otimes $ is the convolution operator.

To determine the formula of \eqref{ZEqnNum378013}, the estimation of the unknown parameters $\omega $ in \eqref{ZEqnNum378013}, $\delta $ in \eqref{ZEqnNum652581}, and $\alpha $, $\gamma $ in  \eqref{ZEqnNum788492}, is as follows. An ${\rm {\mathcal L}}$  Laplace approximation of a marginal likelihood \cite{lapl,nsr2} can be derived to evaluate the estimations of the unknown parameters at a particular $\vec{R} \left(N\right)$ (see \eqref{ZEqnNum227438}), as
\begin{equation} \label{9)} 
{\rm {\mathcal L}}\left(\vec{R} \left(N\right)\left|\alpha ,\gamma ,\delta \right. \right)=F_{{\rm {\mathcal L}}} \left(\vec{R} \left(N\right)\right)\sqrt{\frac{\left(2\pi \right)^{\Omega _{{\rm {\mathcal L}}} } }{\det \left(\Upsilon \right)} },  
\end{equation} 
where $F_{{\rm {\mathcal L}}} \left(\vec{R} \left(N\right)\right)$ is a probability function, as
\begin{equation} \label{10)} 
F_{{\rm {\mathcal L}}} \left(\vec{R} \left(N\right)\right)=\Pr \left(\vec{R} \left(N\right),\alpha ,\gamma ,\delta ,\Lambda \left(\vec{Q}^{tot} \left(N\right)\right)\right), 
\end{equation} 
where $\Lambda \left(\vec{Q}^{tot} \left(N\right)\right)\in {\rm {\mathbb{R}}}^{\Omega _{{\rm {\mathcal L}}} } $ is an approximation of $\vec{Q}^{tot} \left(N\right)$, $\Omega _{{\rm {\mathcal L}}} $ is the order of approximation, while $\Upsilon $ is defined as
\begin{equation} \label{11)} 
\Upsilon ={\rm {\mathcal H}}^{-1} \left(-\log F_{{\rm {\mathcal L}}} \left(\vec{R} \left(N\right)\right)\right), 
\end{equation} 
where ${\rm {\mathcal H}}^{-1} $ is the inverse of a Hessian ${\rm {\mathcal H}}$. 

As follows, the unknown parameters can be evaluated from the noisy empirical vector \eqref{ZEqnNum227438}, therefore the $\left\langle Q^{tot} \left(N\right)\right\rangle $ estimate of $Q^{tot} \left(N\right)$ can be determined via the formula of \eqref{ZEqnNum378013}.

\subsection{Entropy Rate on a Lie-Group}
The entropy rate \cite{erate} in the protocol can be formalized using Lie algebra theory \cite{lie0,lie1,lie2}, in the following manner.

At a given $Q\left(\Phi ^{c} \left(f\right)\right)$ at a particular failure domain ${\rm {\mathcal R}}_{f} $, let
\begin{equation} \label{ZEqnNum284618} 
G_{f} \equiv G\left({\rm {\mathcal R}}_{f} ,Q\left(\Phi ^{c} \left(f\right)\right),c\right)\in SE\left(n\right) 
\end{equation} 
be a group function on the $n=2$ dimensional Lie group $SE\left(n\right)=SE\left(2\right)$, defined as   
\begin{equation} \label{2)} 
G_{f} =\exp \left({\rm {\mathcal R}}_{f} X_{1} +Q\left(\Phi ^{c} \left(f\right)\right)X_{2} \right)\exp \left(c\cdot X_{3} \right), 
\end{equation} 
where $c$ is a constant set to $c=0$, while $X_{1} ,X_{2} $ and $X_{3} $ are basis matrices for the Lie algebra \cite{lie1,lie2} $SE\left(2\right)$, as
\begin{equation} \label{3)} 
{{X}_{1}}=\left( \begin{matrix}
   0 & 0 & 1  \\
   0 & 0 & 0  \\
   0 & 0 & 0  \\
\end{matrix} \right),{{X}_{2}}=\left( \begin{matrix}
   0 & 0 & 0  \\
   0 & 0 & 1  \\
   0 & 0 & 0  \\
\end{matrix} \right),{{X}_{3}}=\left( \begin{matrix}
   0 & -1 & 0  \\
   1 & 0 & 0  \\
   0 & 0 & 0  \\
\end{matrix} \right).
\end{equation} 
Then, let 
\begin{equation} \label{xZEqnNum378013} 
\varphi _{f} \equiv \varphi \left(G\left({\rm {\mathcal R}}_{f} ,Q\left(\Phi ^{c} \left(f\right)\right),c\right),f\right) 
\end{equation} 
be a PDF that characterizes the distribution of the group function $G_{f} $ at a given $f$. 

For \eqref{xZEqnNum378013}, the Lie derivative $X'_{i} \varphi _{f} $, $i=1,2,3$,  is defined as
\begin{equation} \label{5)} 
X'_{i} \varphi _{f} =\left[\frac{d}{df} \varphi \left(G_{f} \circ e^{fX_{i} } \right)\right]_{f=0} , 
\end{equation} 
where $\varphi \left(G_{f} \circ e^{fX_{i} } \right)$ is a PDF of $\left(G_{f} \circ e^{fX_{i} } \right)$, $e^{fX_{i} } $ is a matrix exponential, and $\circ $ is the matrix multiplication operator.

Then, the $S\left(\varphi _{f} \right)$ entropy rate at \eqref{xZEqnNum378013} on a Lie group $SE\left(2\right)$ is yielded as
\begin{equation} \label{ZEqnNum619534} 
S\left(\varphi _{f} \right)=-\int\limits _{SE\left(2\right)}\varphi _{f} \left(G_{f} \right)\log \varphi _{f} \left(G_{f} \right)dG_{f}  , 
\end{equation} 
while the $S'\left(\varphi \right)$ change of the entropy rate of \eqref{ZEqnNum619534} is as 
\begin{equation} \label{7)} 
\begin{split}
   {S}'\left( {{\varphi }_{f}} \right)&=\frac{dS\left( {{\varphi }_{f}} \right)}{df} \\ 
 & =-\int\limits_{SE\left( 2 \right)}{\left( \frac{\partial {{\varphi }_{f}}}{\partial f}\log {{\varphi }_{f}}+\frac{\partial {{\varphi }_{f}}}{\partial f} \right)d{{G}_{f}}}.  
\end{split}
\end{equation} 
Applying the derivations for the $m$ failure domains ${\rm {\mathcal R}}_{f} $ , $f=1,\ldots ,m$, the $S_{\Sigma } \left(\varphi \right)$ total entropy rate is 
\begin{equation} \label{8)} 
S_{\Sigma } \left(\varphi \right)=-\int\limits _{1}^{m}\int\limits _{SE\left(2\right)}\varphi _{f} \left(G_{f} \right)\log \varphi _{f} \left(G_{f} \right)dG_{f} df  ,   
\end{equation} 
while $S'_{\Sigma } \left(\varphi \right)$ the derivative of $S_{\Sigma } \left(\varphi \right)$ is as
\begin{equation} \label{9)} 
S'_{\Sigma } \left(\varphi \right)=-\int\limits _{1}^{m}\int\limits _{SE\left(2\right)}\left(\frac{\partial \varphi _{f} }{\partial f} \log \varphi _{f} +\frac{\partial \varphi _{f} }{\partial f} \right)dG_{f} df  . 
\end{equation} 

\section{Numerical Evaluation}
\label{nume}
The numerical evaluation serves illustration purposes in random quantum network settings. As future work, our aim is to utilize an advanced network simulation framework \cite{sim}.  

\subsection{CP-EAR and PR-EAR}
In this subsection, the CP-EAR and PR-EAR coefficients are illustrated. 

The analysis assumes $f=1,\ldots ,100$ failure domains in random quantum network scenarios $N_{s} $, $s=1,2$, such that distribution of $\Pr \left(f\right)$-s are drawn from a ${\rm {\mathcal U}}$ uniform distribution, $\left\{\Pr \left(f\right)\right\}_{f=1}^{100} \in {\rm {\mathcal U}}$.

\begin{center}
\begin{figure*}[!h]
\begin{center}
\includegraphics[angle = 0,width=1\linewidth]{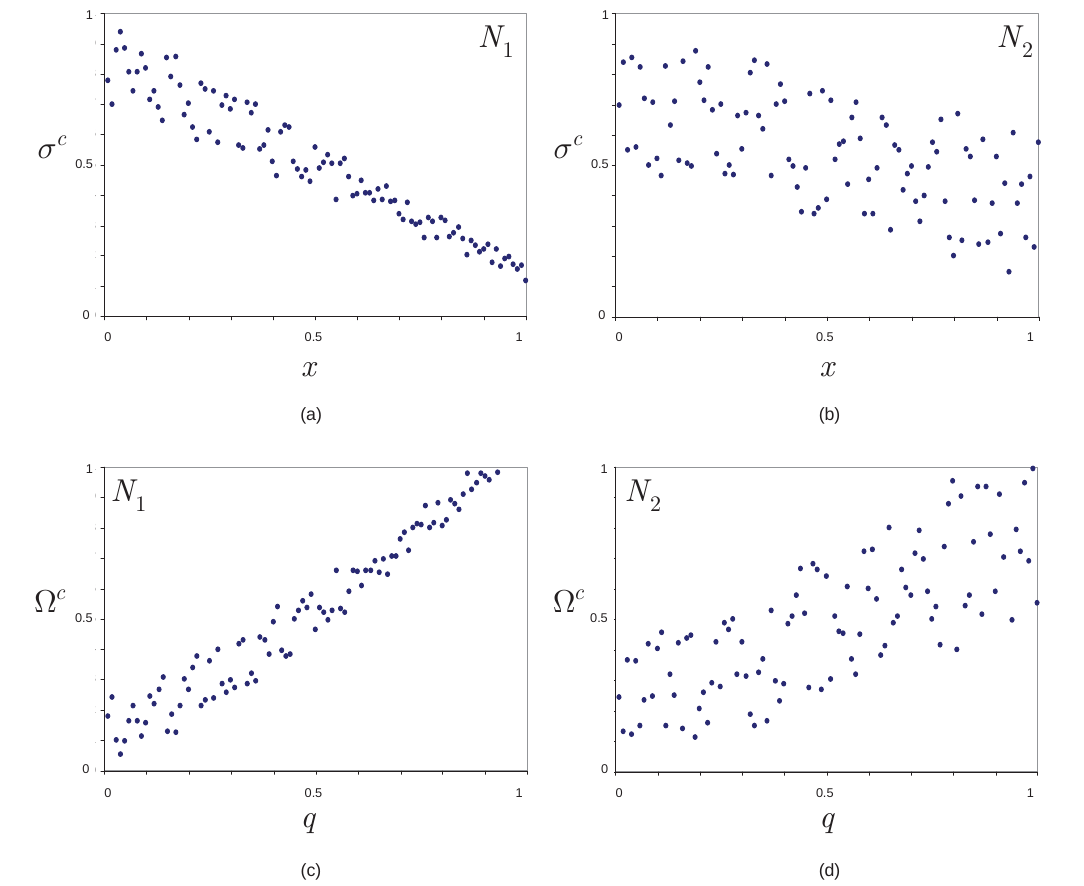}
\caption{The CP-EAR coefficient (a)-(b), and the PR-EAR coefficient (c)-(d). (a) A distribution of the $\sigma ^{c} \left(\Phi ^{c} \left(f\right)\right)$ coefficient for a random network scenario $N_{s} $, $s=1$, in function of $x$, $\Phi ^{c} \left(f\right)\ge x$, $f=1,\ldots ,100$. (b) A distribution of the $\sigma ^{c} \left(\Phi ^{c} \left(f\right)\right)$ coefficient for a random network scenario $N_{s} $, $s=2$, in function of $x$, $\Phi ^{c} \left(f\right)\ge x$, $f=1,\ldots ,100$. (c) Distribution of  $\Omega ^{c} \left(\Phi ^{c} \left(f\right)\right)$ in function of $q$, $q=\Pr \left(f\right)$, for $N_{1} $. (d) Distribution of  $\Omega ^{c} \left(\Phi ^{c} \left(f\right)\right)$ in function of $q$, $q=\Pr \left(f\right)$, for $N_{2} $.} 
 \label{figA1}
 \end{center}
\end{figure*}
\end{center}

The distributions of the $\sigma ^{c} \left(\Phi ^{c} \left(f\right)\right)$ coefficient for random quantum network scenarios $N_{s} $, $s=1,2$, in function of $x$, $\Phi ^{c} \left(f\right)\ge x$, are depicted in \fref{figA1}(a)-(b). The corresponding $\Omega ^{c} \left(\Phi ^{c} \left(f\right)\right)$ values of $N_{s} $, $s=1,2$, in function of $q$, $q=\Pr \left(f\right)$, are depicted in \fref{figA1}.(c)-(d).

\subsection{DD-EAR}
In this subsection, the DD-EAR coefficient is illustrated for random quantum network scenarios $N_{s} $, $s=1,2$, with $f=1,\ldots ,100$.

\begin{center}
\begin{figure*}[!h]
\begin{center}
\includegraphics[angle = 0,width=1\linewidth]{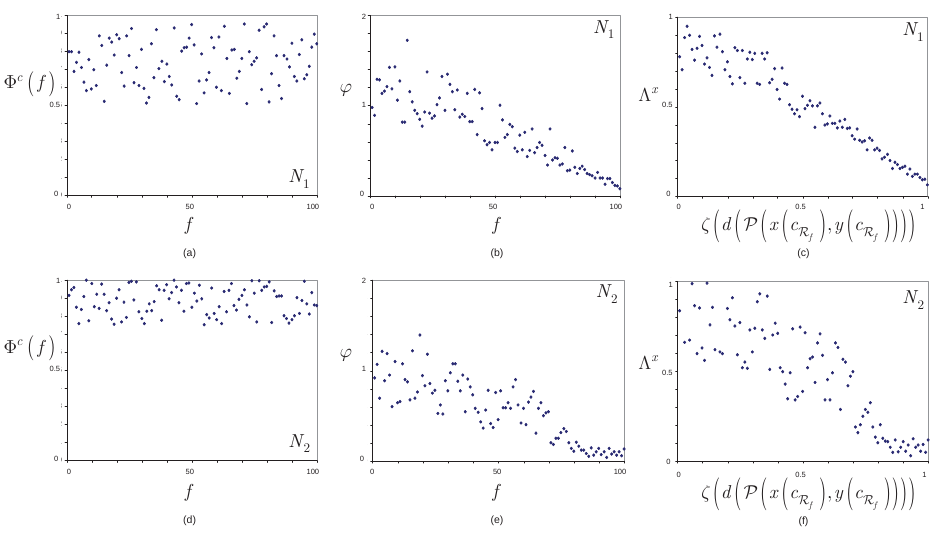}
\caption{The DD-EAR coefficient. The distribution of $\Phi ^{c} \left(f\right)$, $\varphi \left(\Phi ^{c} \left(f\right),r_{{\rm {\mathcal R}}_{f} } \right)$, and $\Lambda ^{x} \left(r_{{\rm {\mathcal R}}_{f} } \right)$, at random quantum network scenarios $N_{s} $, $s=1,2$, with $f=1,\ldots ,100$. (a) A distribution of $\varphi \left(\Phi ^{c} \left(f\right),r_{{\rm {\mathcal R}}_{f} } \right)$ at $N_{1} $, $f=1,\ldots ,100$. (b) A distribution of $\Phi ^{c} \left(f\right)$ at $N_{1} $, $f=1,\ldots ,100$. (c) Distribution of $\Lambda ^{x} \left(r_{{\rm {\mathcal R}}_{f} } \right)$ in function of a normalized hop-distance $\zeta \left(d\left({\rm {\mathcal P}}\left(x\left(c_{{\rm {\mathcal R}}_{f} } \right),y\left(c_{{\rm {\mathcal R}}_{f} } \right)\right)\right)\right)$ at $N_{1} $, $f=1,\ldots ,100$. (d) A distribution of $\varphi \left(\Phi ^{c} \left(f\right),r_{{\rm {\mathcal R}}_{f} } \right)$ at $N_{2} $, $f=1,\ldots ,100$. (e) A distribution of $\Phi ^{c} \left(f\right)$ at $N_{2} $, $f=1,\ldots ,100$. (f) Distribution of $\Lambda ^{x} \left(r_{{\rm {\mathcal R}}_{f} } \right)$ in function of a normalized hop-distance $\zeta \left(d\left({\rm {\mathcal P}}\left(x\left(c_{{\rm {\mathcal R}}_{f} } \right),y\left(c_{{\rm {\mathcal R}}_{f} } \right)\right)\right)\right)$ at $N_{2} $, $f=1,\ldots ,100$.} 
 \label{figA2}
 \end{center}
\end{figure*}
\end{center}

The distribution of the $\Phi ^{c} \left(f\right)$ and $\varphi \left(\Phi ^{c} \left(f\right),r_{{\rm {\mathcal R}}_{f} } \right)$ coefficients of  $\Lambda ^{x} \left(r_{{\rm {\mathcal R}}_{f} } \right)$, and the resulting $\Lambda ^{x} \left(r_{{\rm {\mathcal R}}_{f} } \right)$ in function of the normalized hop-distance 
$0\le \zeta \left(d\left({\rm {\mathcal P}}\left(x\left(c_{{\rm {\mathcal R}}_{f} } \right),y\left(c_{{\rm {\mathcal R}}_{f} } \right)\right)\right)\right)\le 1$,
\begin{equation}
\zeta \left( d\left( \mathcal{P}\left( x\left( {{c}_{{{\mathcal{R}}_{f}}}} \right),y\left( {{c}_{{{\mathcal{R}}_{f}}}} \right) \right) \right) \right)=\tfrac{d\left( \mathcal{P}\left( x\left( {{c}_{{{\mathcal{R}}_{f}}}} \right),y\left( {{c}_{{{\mathcal{R}}_{f}}}} \right) \right) \right)}{d\left( {{\mathcal{P}}^{*}}\left( x\left( {{c}_{{{\mathcal{R}}_{f}}}} \right),y\left( {{c}_{{{\mathcal{R}}_{f}}}} \right) \right) \right)},
\end{equation}
where ${\rm {\mathcal P}}\left(x\left(c_{{\rm {\mathcal R}}_{f} } \right),y\left(c_{{\rm {\mathcal R}}_{f} } \right)\right)$ is a shortest entangled path between $x\left(c_{{\rm {\mathcal R}}_{f} } \right)$ and $y\left(c_{{\rm {\mathcal R}}_{f} } \right)$ in ${\rm {\mathcal R}}_{f} $, while $d\left({\rm {\mathcal P}}^{*} \left(x\left(c_{{\rm {\mathcal R}}_{f} } \right),y\left(c_{{\rm {\mathcal R}}_{f} } \right)\right)\right)$ is an upper bound on $d\left({\rm {\mathcal P}}\left(x\left(c_{{\rm {\mathcal R}}_{f} } \right),y\left(c_{{\rm {\mathcal R}}_{f} } \right)\right)\right)$ in ${\rm {\mathcal R}}_{f} $, for random quantum network scenarios $N_{s} $, $s=1,2$ are depicted in \fref{figA2}. 

\section{Conclusions}
\label{sec5}
Here, we defined entanglement accessibility measures to evaluate the ratio of accessible quantum entanglement at complex failure events in the quantum Internet. A complex failure is modeled by a complex failure domain, which identifies a set of quantum nodes and entangled connections affected by that failure. We introduced the terms entanglement accessibility ratio and occurrence coefficient to characterize the availability of entanglement in a multiple failure setting. We proposed an algorithm to derive the occurrence coefficient via an empirical estimation observable from the evaluated parameters of the analyzed quantum network. The defined metrics and algorithm can be applied efficiently in experimental quantum Internet scenarios.


\section*{Acknowledgements}
The research reported in this paper has been supported by the Hungarian Academy of Sciences (MTA Premium Postdoctoral Research Program 2019), by the National Research, Development and Innovation Fund (TUDFO/51757/2019-ITM, Thematic Excellence Program), by the National Research Development and Innovation Office of Hungary (Project No. 2017-1.2.1-NKP-2017-00001), by the Hungarian Scientific Research Fund - OTKA K-112125 and in part by the BME Artificial Intelligence FIKP grant of EMMI (Budapest University of Technology, BME FIKP-MI/SC).

\newpage
\appendix
\setcounter{table}{0}
\setcounter{figure}{0}
\setcounter{equation}{0}
\setcounter{algocf}{0}
\renewcommand{\thetable}{\Alph{section}.\arabic{table}}
\renewcommand{\thefigure}{\Alph{section}.\arabic{figure}}
\renewcommand{\theequation}{\Alph{section}.\arabic{equation}}
\renewcommand{\thealgocf}{\Alph{section}.\arabic{algocf}}

\setlength{\arrayrulewidth}{0.1mm}
\setlength{\tabcolsep}{5pt}
\renewcommand{\arraystretch}{1.5}
\section{Appendix}
\subsection{Abbreviations}
\begin{description}
\item[API] Application Programming Interface
\item[CP-EAR] Cumulative Probability of Entanglement Accessibility Ratio
\item[EAR] Entanglement Accessibility Ratio
\item[O-EAR] Occurrence of Entanglement Accessibility Ratio
\item[PDF] Probability Density Function
\item[PR-EAR] Probabilistic Reduction of Entanglement Accessibility Ratio
\item[QKD] Quantum Key Distribution
\item[DD-EAR] Domain Dependent Entanglement Accessibility Ratio

\end{description}

\end{document}